# Plasmofluidic Single-Molecule Surface Enhanced Raman Scattering from Dynamic Assembly of Plasmonic Nanoparticles

Partha Pratim Patra, Rohit Chikkaraddy, Ravi P. N.Tripathi, Arindam Dasgupta, G.V. Pavan Kumar*

Photonics & Optical Nanoscopy Laboratory, h-cross,

Indian Institute of Science Education and Research, Pune – 411008, INDIA

*corresponding author's e-mail: pavan@iiserpune.ac.in

**Abstract**

Single-molecule surface enhanced Raman scattering (SM-SERS) is one of the vital applications of plasmonic nanoparticles. The SM-SERS sensitivity critically depends on plasmonic hot-spots created at the vicinity of such nanoparticles. In conventional fluid-phase SM-SERS experiments, plasmonic hot-spots are facilitated by chemical aggregation of nanoparticles. Such aggregation is usually irreversible, and hence, nanoparticles cannot be re-dispersed in the fluid for further use. Here, we show how to combine SM-SERS with plasmon-polariton assisted, reversible assembly of plasmonic nanoparticles at an unstructured metal-fluid interface. One of the unique features of our method is that we use a single evanescent-wave optical excitation for nanoparticle-assembly, manipulation and SM-SERS measurements. Furthermore, by utilizing dual excitation of plasmons at metal-fluid interface, we create interacting assemblies of metal nanoparticles, which may be further harnessed in dynamic lithography of dispersed nanostructures. Our work will have implications in realizing optically addressable, plasmofluidic, single-molecule detection platforms.

Interaction of light with nanoscale matter at sub-wavelength regimes has gained attention in various branches of science and technology. The significance of this interaction is driven by the fact that quantum systems such as atoms, molecules, quantum dots and their varieties can be individually coupled to electromagnetic fields at the nanoscale.[1-5] Studying molecules at single-copy-limit not only has tremendous implications in chemical and biological sensing,[2, 6, 7] but also can lead to new physical insights about molecules that are not evident when their respective ensembles are interrogated.[8-10] Controlling electromagnetic fields at nanoscale is one of the important tasks of nanophotonics, and to serve this purpose - plasmons, which are charge density oscillations at metal dielectric interface, have been utilized extensively.[11-13] Plasmons are usually produced in two major varieties: one is the localized surface plasmons (LSPs) facilitated by metal nanoparticles and junctions between nanostructures[14], and the other is the surface plasmon polaritons (SPPs) that are usually propagating in nature, and can be produced on a variety of platforms such as films, nanowires etc.[15]

There are various applications of plasmons[16-20] of which two are directly relevant to this study: solution-phase single molecule surface enhanced Raman scattering (SM-SERS) and trapping of nano-scale objects in plasmofluidic field (i.e., plasmonic field in fluidic environment). The SM-SERS was first achieved in 1997,[21, 22] and ever since it has emerged as one of the important nanoscale spectroscopy techniques to interrogate both molecules and localized optical fields.[23-29] Plasmon-assisted optical trapping[30-34] can be utilized to trap and manipulate micro and nanoscale objects at metal-fluid interface[35-41] with laser powers significantly lower than the conventional optical trapping.

One of the objectives of this work was to integrate SM-SERS with plasmon-assisted assembly of nanoparticles in solution. This integration is advantageous, especially in the context of optofluidics, for the following reason. In most of the solution-phase SM-SERS experiments [23, 42-44] the plasmonic hot-spots are facilitated by adding chemical aggregating agent such as NaCl or KCl. These chemicals agglomerate the nanoparticles irreversibly, and hence the aggregates cannot be further used as uniformly dispersed, plasmonic nanoscale entities. Also, such procedure of aggregation is not conducive for micro/nano-fluidic environments, where a relatively small aggregate can block the flow of the liquid. Therefore, there is an imperative need to temporarily assemble nanoparticles in a controlled way, and re-disperse them back in the fluid after SM-SERS interrogation.

In this paper, we show how to combine single-molecule SERS with plasmon-assisted assembly and manipulation of metal nanoparticles without the use of aggregating agents. Importantly, *a single evanescent-wave optical excitation was used to achieve both plasmonic assembly and single-molecule SERS, simultaneously.* To best of our knowledge, this is the first report on solution-phase SM-SERS experiments on dynamic assembly of metal nanoparticles by evanescent-wave excitation of plasmons. By harnessing the optical fields created by SPPs in a metal film, we create large-area assembly of metal nanoparticles, and further utilize the combination of SPP and LSP fields to capture single-molecule SERS signatures at metal-fluid interface. In addition to this, we show the potential of our method to achieve dynamic, large-area lithography of nanoparticles at metal-fluid interface by employing dual-optical excitation in total-internal reflection geometry.

**Results**

**Combining plasmon-assisted assembly of nanoparticles with confocal SERS microscopy**

We first addressed the issue of how to effectively employ a *single* evanescent-wave optical excitation at metal-fluid interface to *simultaneously* assemble plasmonic nanoparticles and perform Raman scattering. For this, we utilized total-internal reflection (TIR) based plasmonic excitation of an unstructured metal film in Kretschmann geometry. This plasmon-polariton excitation created an optical potential, which was further utilized to assemble and manipulate nanoparticles. Importantly, the same TIR illumination was also utilized for exciting the Raman scattering of the assembled nanoparticles.

Figure 1a shows the optical schematic of the set-up used for plasmon-assisted assembly of metal nanoparticles and SM-SERS measurements. We used a Dove prism (N-BK7, refractive index = 1.519), on to which a silver-coated glass cover-slip was adhered through an optically-matched oil. The thickness of the metal film was 40±5nm. The excitation source was a 532nm, continuous-wave, frequency doubled Nd:YAG laser (power = up to 200mW) which was p-polarized by a λ/2 plate and routed into the prism for total internal reflection by weakly focusing lens L (focal length = 150mm). The angle of incidence at the interface was $76.7^0$ from the normal to the surface, which was close to the TIR angle of Ag-water interface. (see Supplementary Figure-1). The optical image and the Raman scattered signal from the metal-water interface was collected via an objective lens (OBL). For conventional optical imaging, a 4x, 0.13 NA lens was used, whereas for the SM-SERS measurements, a water immersion objective lens (60x, 1.0NA) was used. The captured light was routed towards the camera or into a high resolution, confocal Raman spectrometer (LabRam HR, 789mm focal length, pinhole size = 400µm). The p-polarized Gaussian excitation beam creates an elliptical projection upon total internal reflection at the metallic surface, which was imaged in figure 1c. This particular excitation profile acts as an optical potential, which was harnessed for plasmofluidic trapping and SM-SERS measurements. Upon introducing bimetallic core-shell nanoparticle -solution (80µL) on to the Ag surface, the nanoparticles slowly (within 15min) assembled at the location of the elliptical spot. One such assembly of nanoparticles is shown in figure 1c by filtering the incident laser beam. The open circle shown in figure 1c is the location from where the confocal Raman signals were captured.

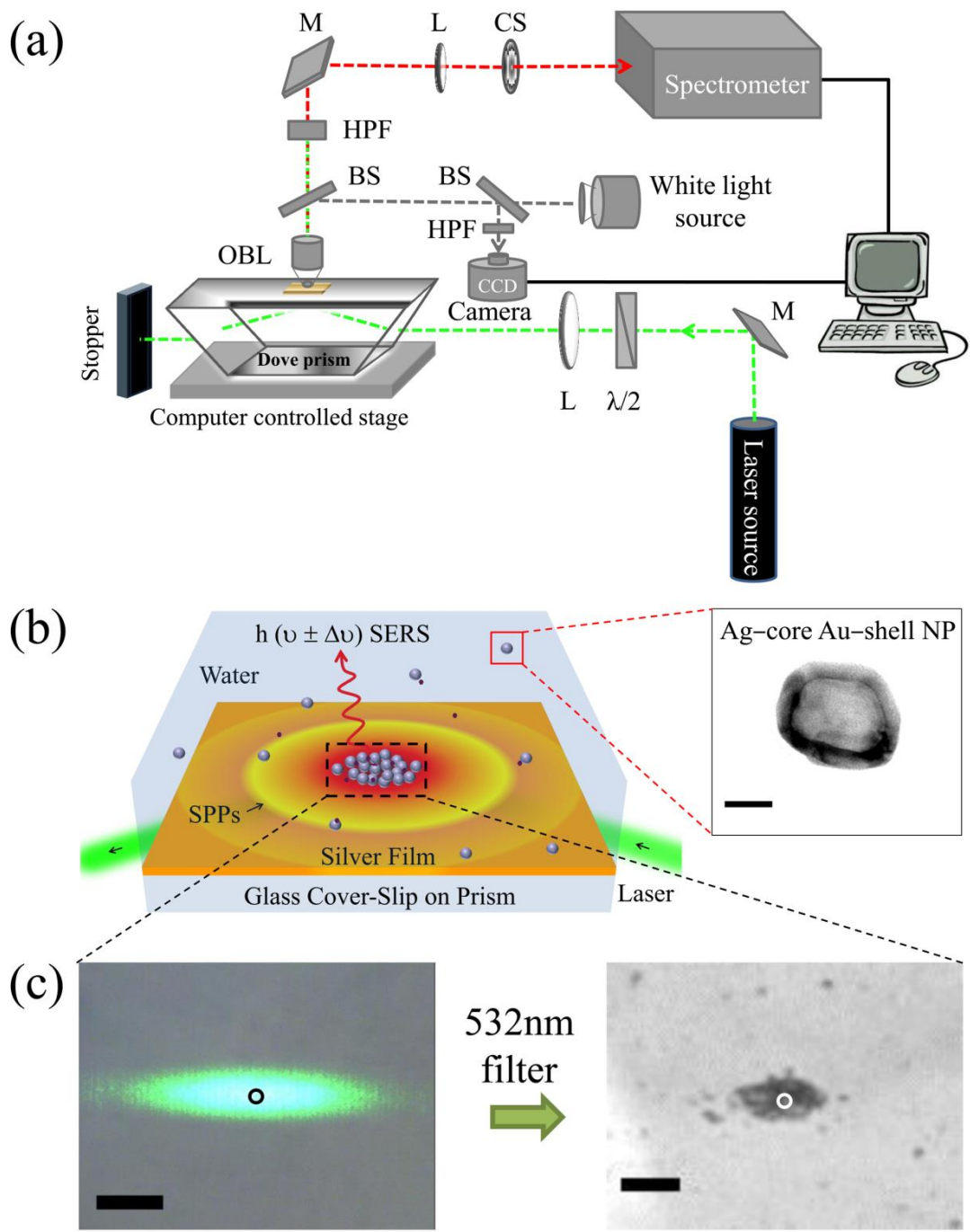

Figure 1. **Concomitant SERS microscopy with plasmon-assisted assembly of nanonnparticles at metal-fluid interface**:(a) Optical set-up of confocal Raman detection microscope coupled with Dove-prism based plasmon excitation. (b) Schematic illustration of the simultaneous plasmonic assembly of nanoparticles and SERS. Black arrows indicate the direction of 532nm laser (green beam) used for SPP excitation of the metal film deposited over glass coveslip. Grey spheres are the metal nanoparticles dispersed in the water medium and red dots indicate the SERS-active molecules. Inset on the right shows the TEM image of Ag core- Au shell nanoparticle used in this experiment. Scale bar is 20 nm (c) On the left is an unfiltered optical bright-field image of 532nm elliptical excitation on the metal film; the right side image shows the plasmon-assisted assembly of nanoparticles in the excitation region (532nm rejected using filter). The open circles in (c) indicates the area from which SERS signals were collected. Scale bar is 80 μm. The acronyms in (a) are M-mirror, HPF-

high pass filter, L-lens, CS-confocal slit, BS-beam splitter, OBL-objective lens, CCD - charged-coupled device, λ/2- half wave plate.

**Choice of the metal film and nanoparticles for our study:**

Trapping of plasmonic nanoparticles on metal film is extremely sensitive to three parameters: plasmon resonance of the nanoparticles, SPP resonance of the film and excitation wavelength (see Supplementary Figure-2 and 3 for the data on resonances of various metal films and nanoparticles). We found that if the resonance of nanoparticles has a large overlap with the excitation wavelength, as in case of Au nanoparticles at 532nm, then nanoparticles usually escaped from the trap, and were not suitable for our experiments. Such propulsion of Au nanoparticles in SPP field has also been reported by various researchers.[45-48] As an alternate to Au nanoparticles, we used Ag-core Au-shell (Ag@Au) nanoparticles. Recently we have shown[44] that Ag@Au nanoparticle architectures are excellent substrates for single-molecule SERS experiments at visible frequencies. There are at least three advantages[49] of using such nanoparticles: 1) they facilitate optical near-fields comparable to Ag nanoparticle and better than Au nanoparticle; 2) by varying the thickness of the gold shell, one can systematically tune the plasmon resonance from visible to infra-red wavelength; 3) since these particles have gold shell, they provide biocompatibility[49] and stability compared to Ag nanoparticles. With this hindsight, we tuned the Au shell-thickness of Ag@Au nanoparticle such that we obtained a plasmon resonance at 490 nm. This value was slightly off-resonance from the 532nm excitation and hence suitable for both plasmonic assembly and SM-SERS experiments. Also, for the Ag film-water interface, the SPR resonance wavelength was found to be 500nm which was again off-resonance from our 532 excitation. Taking all the resonances and other advantages into consideration, we found Ag@Au nanoparticle suitable for both plasmonic assembly and single-molecule SERS experiments.

**Assembly of Ag@Au nanoparticle in plasmofluidic field**

Figure 2a shows the time series images of the assembled Ag@Au nanoparticles on a metal film in presence of p-polarized evanescent-wave excitation (see Supplementary Movie 1 and 2). We observed a gradual aggregation of nanoparticle spanning over a time period of 30 minute. Between 06 minute to 16 minute, there was a steady increase in the number of nanoparticles assembled in the illuminated spot, but after 20 minute, we found that the number of nanoparticle in the illuminated region reached a saturation point in time, and thereafter the aggregate reached a steady state.

Figure 2b shows the aggregate-growth kinetics of the nanoparticles in the dotted ellipse region. We captured the scattering intensity coming out of the elliptical region as a function of time, and the points in figure 2b shows the experimental data. The experimental growth kinetics data fitted well to a least-square Boltzmann equation (see Supplementary Note-1 for details). The accumulation of nanoparticles in the illuminated region was mainly due to the combination of fluid convection and SPP forces experienced by the nanoparticles. Similar kind of plasmofluidic assembly of dielectric microspheres on a gold nano-film has been reported by Garcés-Chávez et al.[35]. The excitation of SPP on the film creates a gradient of temperature just above the illuminated region. The temperature distribution in the surrounding water medium is governed by the following equation[50]:

$$\rho c \left[ \partial_t T(r,t) + \nabla \cdot ( T(r,t) \mathsf{v}(r,t)) \right] - \kappa \nabla^2 T(r,t) = 0 \qquad (1)$$

Where $\mathsf{v}(r,t)$ is the fluid velocity and $\nabla \cdot (T\mathsf{v})$ is the nonlinear convective term. $\kappa$, $\rho$, $c$ are the thermal conductivity, mass density, and specific heat capacity of water at constant pressure, respectively. The temperature gradient creates convection in the fluid, and the nanoparticles move from the cooler region of the fluid into the hotter region, i.e. towards the illuminated spot. Furthermore, the same convection current will try to drive the nanoparticles away from the hotter region, but the gradient force of SPP hold them back from escaping out, thereby trapping them at illuminated spot.

It is to be noted that the LSP resonance of plasmonic nanoparticles also plays an important role in this assembly. Trapping was feasible only when LSP resonance wavelength of nanoparticles was far from the excitation wavelength (532nm). Apart from Ag@Au nanoparticles, we tested this hypothesis for Ag nanoparticles (420nm resonance), and found excellent trapping capability. Interestingly, Ag nanoparticles assembled in the trap at a faster rate compared to Ag@Au nanoparticles (See Supplementary Figure-4).

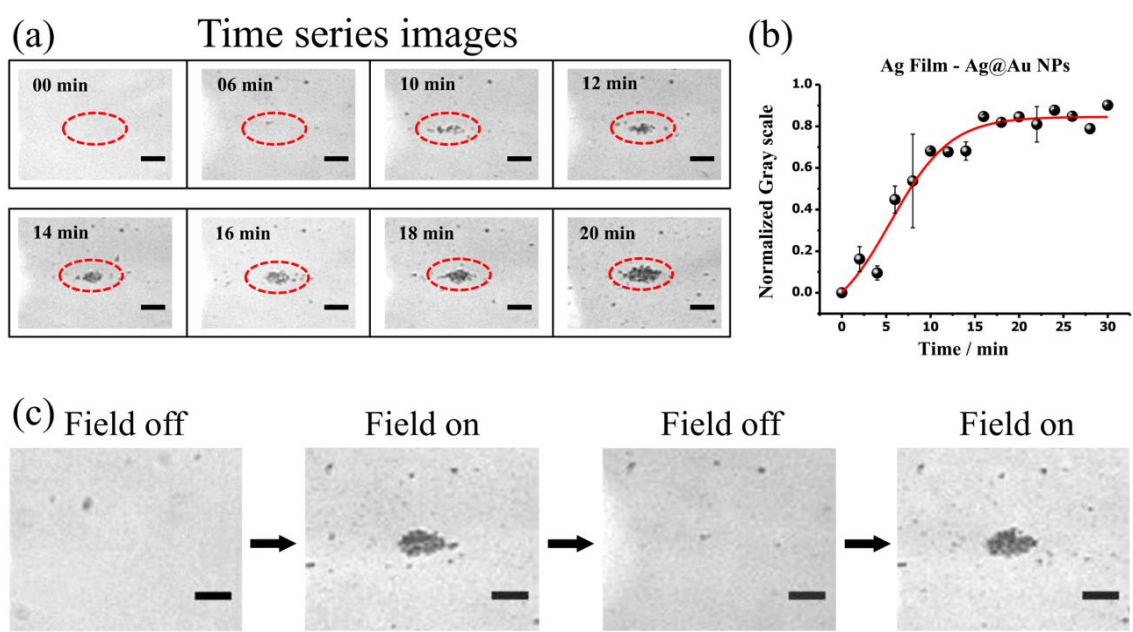

Figure 2. **Plasmon-assisted assembly of nanoparticles at metal-fluid interface: kinetics and repeatability**. (a)Time series images of gradual-accumulation of Ag@Au nanoparticles on silver-film in evanescent optical field indicated by dotted-ellipse in red (see Spplementary Movie 1 and 2). (b) Variation of particle density in the abovementioned field as a function of time, extracted from the images. Solid red-line represents the least-square Boltzmann fit to the experimental data. Error bars on the experimental data points represent the variation (standard deviation) in determining the grey scale values across the optical images obtained throughout the experiment. (c) Optical bright-field images showing capability to repeat the assembly and release of nanoparticles in the presence and absence of plasmonic field at metal-fluid interface. Scale bar is 80 μm for all figures.

Furthermore, we tested the reversibility of the nanoparticle assembly, by switching-off the evanescent-field excitation and later switched it back after a few minutes. Figure 2c shows the optical images of two such cycles of reversible-assembly of Ag@Au nanoparticles where the excitation fields were alternately switched on and off. We tested this process over multiple cycles (at least 5 times) and we found excellent consistency in terms of reversible assembly of nanoparticles at the illuminated location.

**Single-molecule SERS from plasmon-assisted assembly of Ag@Au nanoparticles**

Since many nanoparticles are packed into an assembly on the metal surface, one would expect multiple plasmonic hot-spots in this confinement. Such hot-spots are ideal for SERS experiments.[23, 51] Furthermore, the same optical excitation which is trapping these nanoparticles should also be able excite SERS signatures of the molecules in the hot-spot. Can these hot-spots be sensitive enough to detect single molecule Raman signatures?

To answer this question, we employ

bi-analyte single molecule SERS method developed by Etchegoin and Ru.[42, 52] This technique is based on spectroscopic contrast between two different kinds of SERS-active molecules, and importantly, does not rely on the ratio of number of molecules to the nanoparticles used in the experiments. From a large data set of SERS experiment, reliable single molecule signatures can be extracted by rigorous statistical analysis based on modified principal component analysis (MPCA).[51, 52] The two molecules used for our experiments were nile blue (NB) and rhodamine 6G (R6G) at 5nM concentration. For SM-SERS experiments, we choose 592cm$^{-1}$ and 615cm$^{-1}$ Raman modes of NB and R6G, respectively.

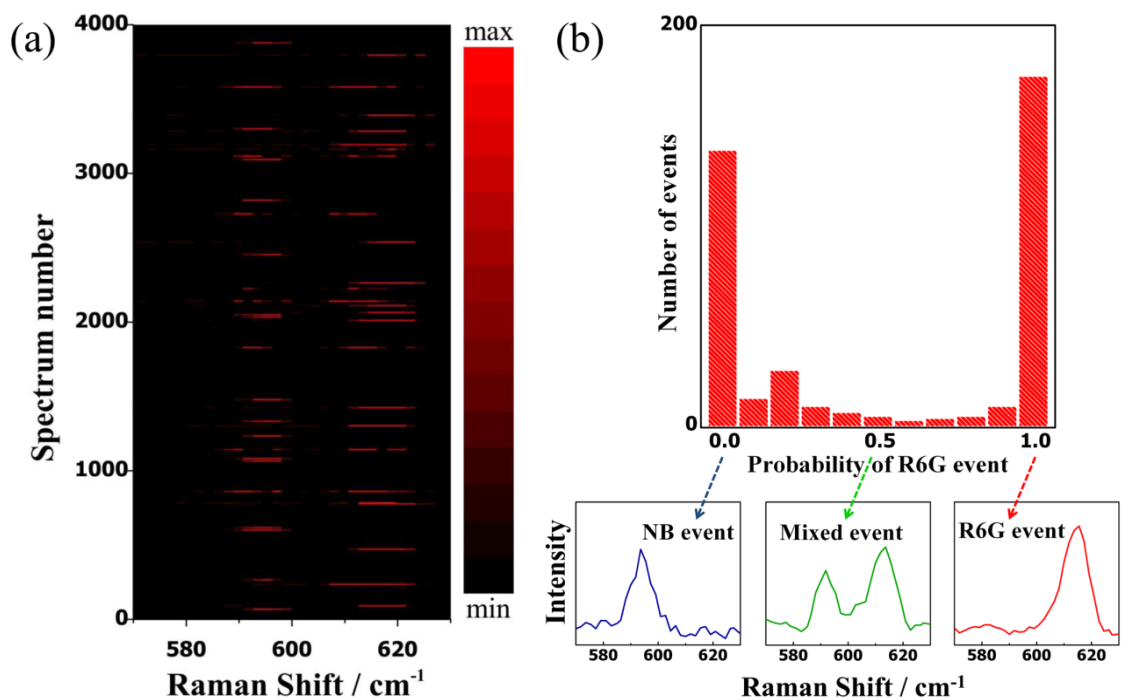

Figure 3. **Single-molecule SERS from plasmon-assisted assembly of Ag@Au nanoparticles on Ag film**. (a) Bi-analyte (nile blue (NB) and rhodamin6G (R6G)) SERS time-series spectra recorded from the open circle region shown in figure 1c. The concentration of both the molecules was 5nM. For modified principal component analysis (MPCA) analysis, 592 $cm^{-1}$ modes of NB and 615 $cm^{-1}$ mode of R6G were considered. 4000 spectra (out of 7000) have been displayed to show spectral fluctuation of single molecules in plasmonic field. Signal acquisition time was 0.1 second. All the data were collected using 532nm evanescent excitation wavelength (20mW power before entering the prism) (b) Single-molecule probability histogram derived from MPCA. Maximum event at 1 and 0 of abscissa confirms single molecule SERS signal from R6G and NB, respectively. The inset shows spectral signatures for 3 different probabilities.

Figure 3a shows time evolution of bi-analyte SM-SERS signal captured from the plasmofluidic trap of core-shell nanoparticles (see the open circle in figure 1c). The total number of spectra collected was 7000, the acquisition time for each spectrum was 0.1s, and the dwell time between consecutive accumulations was 1s. We further processed this data using MPCA[52] to find the principal eigenvalues of the signals generated. All the details of the MPCA analysis can be found in the Supplementary Figure-5 . The results of the MPCA analysis is shown in figure 3b, where the probability of single molecule signature is plotted in the form of a histogram, and accompanying them are the relevant bi-analyte spectra for three different probabilities of single molecule events (0 corresponds to single NB event, 0.5 for mixed event and 1 for single R6G event; the probability for detection of R6G single molecule is slightly greater than NB because electronic resonance of R6G is closer to 532nm excitation). The histogram clearly indicates a large probability of single R6G and single NB molecules' spectral signatures arising from the plasmofluidic trap of nanoparticles. *This data is the central results of this paper*. We emphasize that the single molecule SERS excitation is arising from the same source which is assisting us in trapping the nanoparticles. To best of our knowledge, this is the first report where both plasmofluidic assembly and single-molecule SERS has been concomitantly achieved using a single evanescent-wave excitation. Thus our method not only creates multiple plasmonic hot-spots in a nanoparticle assembly but also probes the single molecule spectral signatures simultaneously.

In our experimental configuration, there are two kinds of plasmonic fields: one is the SPP field created by the metal film and the other is the LSPs of the nanoparticles. Importantly, the SPPs of the film can couple to the LSPs of the nanoparticles.[53-60] This combination of the SPP and LSP can facilitate large electric fields thereby creating plasmonic hot-spots at two locations: junction between film and nanoparticles, and the junction between individual nanoparticles. Furthermore, nanoparticles assemble in both horizontal and vertical direction, leading to multiple hot-spots in the illuminated region of metal-fluid interface. We further corroborated this hypothesis with 3D-finite different time domain calculations (see SupplementaryFigure-6), which indicated near-field enhancement to arise due to film-particle coupling and particle-particle coupling. Such multiple hot-spots at metal-fluid interface can facilitate large scale enhancement, and our experimental result of single molecule spectroscopic detection is one such important consequence due to the near-field coupling between the molecules and nanoparticles.

**Plasmon-assisted manipulation of Ag@Au nanoparticle assembly at metal-fluid interface**

Having shown the capability to perform single-molecule SERS at metal-fluid interface, we asked whether we can systematically manipulate the trapped assembly of nanoparticles, such that they can be utilized as optically addressable, single-molecule SERS substrate in a fluid. To answer this, we probed the effect of moving the elliptical excitation beam on the plasmonic nanoparticle assembly. Figure 4 shows the optical images (time series) of plasmonic nanoparticle transport at metal-fluid interface (seeSupplementary Movie-3). The different locations – P, Q and R in figure 4 indicate the three parking points of elliptical excitation beam. We first started with a plasmonic assembly at P. Next we moved the excitation beam to the location Q, and after a few seconds we observed gradual migration of the plasmonic nanoparticles into new location, Q. About 40 seconds later, we changed the location of optical excitation from Q to R, and within a few seconds, the nanoparticle assembly followed the excitation beam. This experiment revealed two important aspects of the trap: a) capability to accurately manipulate a large assembly of nanoparticles in fluid by plasmon-assisted excitation *without structuring the metal surface*, and b) the robustness to overcome any Brownian drag forces during the movement of the assembly.  These advantages can be further harnessed in plasmon-assisted micro- and nano-fluidics with an added capability of single molecule detection sensitivity.

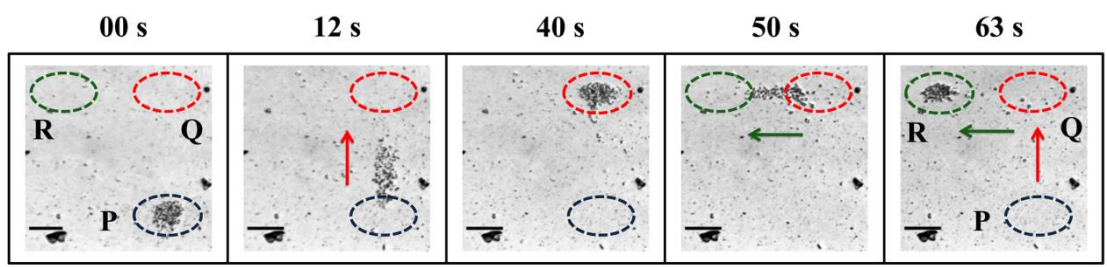

Figure 4. **Plasmon-assisted manipulation of Ag@Au nanoparticle assembly at metal-fluid interface.** Time series images showing transportation of nanoparticle assembly from initially trapped area P to final area R via point Q (see Supplementary Movie-3). The manipulation involves a right-angled turn which was achieved by careful movement of evanescent excitation beam.  The assembly of nanoparticles were accurately transported over a distance of 300 microns within few seconds. Note that the manipulation was achieved on an unstructured metal-fluid interface. Scale bar is 80 μm

It is to be noted that our methodology is versatile and can be applied to different combinations of metal films and plasmonic nanoparticles. Table 1 compares a variety of such combinations for 532nm excitation. Parameters such as nanoparticle assembly time, assembly-size and the SERS signal retrieved from such assemblies are compared. Specifically for SM-SERS studies, we found Ag@Au

nanoparticles and Ag nanoparticles on Ag film to be suitable. Also, other combinations of film and particles can be harnessed for conventional SERS studies.

**Table 1**: Comparison between 6 different combinations of metal film and plasmonic nanoparticles tested for 532nm excitation.

| 532nm evanescent field excitation | 1 | 2 | 3 | 4 | 5 | 6 |
|---|---|---|---|---|---|---|
| | Silver-film + Ag NP | Silver-film + Ag@Au NP | Silver-film + Au NP | Gold-film + Ag NP | Gold-film + Ag@Au NP | Gold-film + Au NP |
| Assembly time | fast (10 min) | Moderate (17 min) | no assembly | fastest (7 min) | moderate (15 min) | no assembly |
| Assembly size | large (400 μm) | Moderate (60 μm) | no assembly | large (500 μm) | moderate (70 μm) | no assembly |
| SERS signal (1s accumulation) | intense | intense (5000 counts) | lowest | moderate | moderate | no signal |

**Towards dynamic, large-area lithography of nanoparticles at metal-fluid interface**

Can we utilize our method in dynamic lithography of dispersed nanoparticles at metal-fluid interface, such that they can be temporarily assembled in a specific geometry and further re-dispersed in the fluid? To answer this question we performed experiments with Ag nanoparticle on Au film (combination 4 in Table 1) with a dual 532nm evanescent-wave excitation in counter-propagating geometry. Figure 5a shows the optical schematic of dual-excitation trap. Two beams were introduced into the Dove prism from opposite ends. The optical image of the two elliptical excitations in the same line is shown in figure 5b, where the distance between the centres of two beams were approximately 340 μm. These excitations further created assemblies of Ag nanoparticles at metal-fluid interface (see Supplementary Movie-4). Interestingly, when these two assemblies were in close proximity, we found interaction between assemblies leading to a linear chain of nanoparticles between them. Figure 5c shows optical image (after rejecting the laser light) of the two plasmonic assemblies, with a thin line of Ag nanoparticles between them. We also observed similar kind of Ag nanoparticle linear-chains when two assemblies were created in an anti-parallel geometry (see figure 5d). The resulting nanoparticle assembly is shown in figure 5e. Since the assemblies are now closer (compared to the case shown in figure 5 c), the nanoparticle bridge formed is thicker in size. It is to be noted that all this manipulation discussed herein is at the metal-fluid interface. If the excitation beam was switched-off, the entire nanoparticle assembly re-dispersed into the solution. This method is reversible over multiple cycles, and can be an excellent preamble to create dynamic assembly of nanoparticles at metal-fluid interface, and can be further harnessed for large-area, dynamic, optical lithography of nanoscale objects.

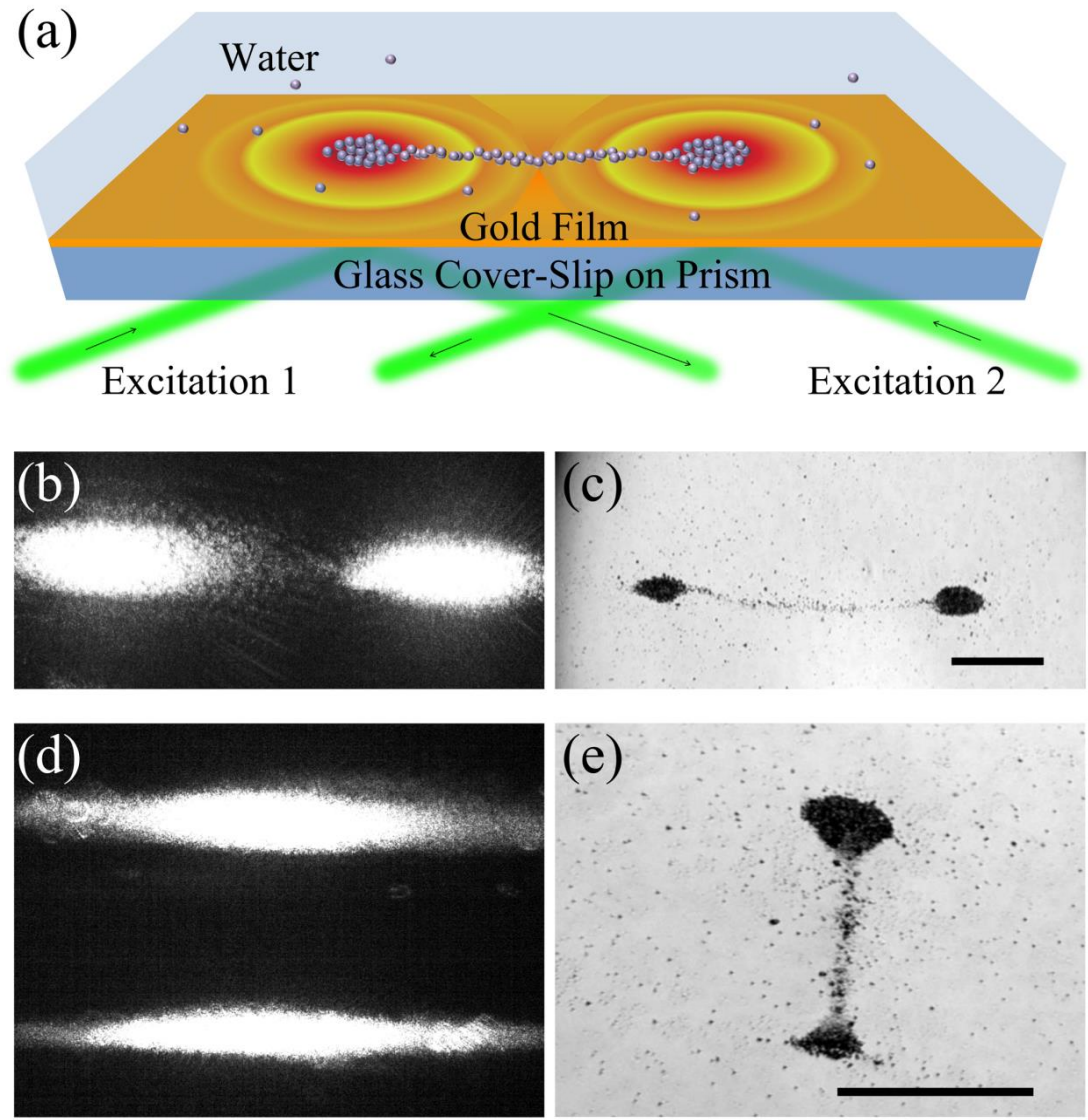

Figure 5. **Interaction between dynamic assemblies of plasmonic nanoparticles at metal-fluid interface**. (a) Schematic representation of dual assembly of Ag nanoparticles on Au film coupled to a prism: green lines are the two laser beams of 532nm introduced at two opposite ends of the prism. Black arrows indicate the direction of the laser beams (see Supplementary Movie-4). Optical images of dual laser excitation in (b) collinear geometry with opposite k-vectors and (d) anti-parallel geometries with opposite k-vectors. (c) and (e) are optical images (532nm-rejected) of Ag nanoparticles assembly due to excitation in shown in (b) and (d), respectively. A thin line of nanoparticles bridging the two assemblies are evident. Note that all these assemblies are at metal-

fluid interface, and hence can be further manipulated to achieve dynamic lithography. Scale bar = 100 μm for both (c) and (e).

**Discussion**

We have used an unstructured metal film and plasmonic nanoparticles to form a sensitive molecular detection platform. The sample preparation was easy to reproduce, and can be readily integrated with any microfluidic device. There are situations in microfluidics where one needs to control mixing of molecules and/or nanoparticles for a brief period of time. By using our method, one can assemble nanoparticles for a short instance of time, spectrally probe a mixing reaction, and release them back into the solution. This capability to trap, probe and release nanoparticles in a controlled way, especially using optical means, will have implications in driving optofluidic chemical and biochemical reactors. It is to be noted that our methodology is not restricted to metal nanoparticles alone. One can trap two different kinds of samples, such as biological cells and nanoparticles in separate traps, and bring them together in close proximity to obtain a mixture. Since assembly and probing is done through optical means, one can also utilize the same optical fields to catalyze photochemical reactions.

To conclude, we show how plasmon-assisted assembly and single-molecule SERS can be achieved with a single evanescent-wave excitation on an unstructured metal film. Such dual-function of plasmons at metal-fluid interface can have relevance in trace detection of molecules at interface without having to structure the metal surface. Furthermore, on the same platform, we have shown how nanoparticles can be reversibly assembled and manipulated with ease. This is also unique method to prepare and optically manipulate SERS substrates at metal-fluid interface that is sensitive enough to detect single-molecule spectroscopic signatures. Interestingly, when we used two excitation beams in our experiments, we could create interacting nanoparticle assemblies. The cross-talk between the assemblies led to the formation of a nanoparticle bridge between the optical potentials paving a new way to assemble nanostructure at metal-fluid interface. We envisage that interacting optical potential can be further harnessed in large-area dynamic lithography of nanoparticles at metal-fluid interface.

Such capability will of course find tremendous use in micro- and nano-plasmofluidic applications, where plasmonic fields can drive, trap and probe nanoscale objects.

**Methods:**

**Preparation of Metal Thin Films:**

Glass cover slip was washed by acetone followed by drying was used for deposition of metal films. DC sputtering technique was employed to deposit 40nm thick gold or silver on glass-cover slip at a rate of 0.1Å/s in argon atmosphere.

**Preparation of metallic nanoparticles:**

Silver and gold nanoparticles were synthesized by citrate reduction procedure.[61] Ag nanoparticles: 250 mL, 5.3 × $10^{-4}$ M aqueous solution of $AgNO_3$ was heated and brought to boiling using a magnetic stirrer cum hot plate. Then 5mL of 1% sodium citrate solution was added while stirring the solution and kept for ~1 hour.

Au nanoparticles: 250 mL, 6.7 × $10^{-4}$ M aqueous solution of $HAuCl_4$ was brought to boiling using magnetic stirrer cum hot plate. Then 20 mL of 1% sodium citrate solution was added while stirring the solution and kept for ~1 hour.

Ag@Au nanoparticles: Ag@Au nanoparticles were prepared by seed-growth method,[44] using Ag colloids as seeds. 12.5 mL Ag colloid solution was diluted with 10 mL of mili-Q water, and 2.5 mL of 6.25 × $10^{-3}$ M $NH_2OH$, HCl and 2.5mL of $4.65\times10^{-4}$ M $HAuCl_4$ were added drop wise (ca. 2 mL/min) by two separate pipettes with vigorous stirring. The amount of $HAuCl_4$, here, governs the Au to Ag ratio. The stirring was continued for 45 min to complete reduction. The Ag@Au nanoparticles were characterized by UV-Vis spectrometer and transmission electron microscope (FEI Tecnai F20, 200kV).

**Sample preparation for SM -SERS experiments:**

Ag@Au nanoparticles were mixed with two dye molecules, nile blue chloride (NB) and rhodamine 6G chloride (R6G) such that the final concentration of dye molecules is 5nM and 80 µL of such solution was drop casted on glass cover-slip pre-deposited with 40nm thick silver film. Series of SERS spectrum were collected with the acquisition time for each spectrum was 0.1 s and the refresh time for consecutive two spectra was 1 s. The laser power used was 20 mW before entering the prism. The signal collection spot was fixed to 5 µm by keeping the confocal hole size as 400 µm. Water immersion 60x objective lens (NA=1) was used for signal collection. *Note that there were no external aggregating agents used.* For concentration dependent SERS measurements, all the parameters of the experiments were held same as single molecule SERS studies except the molecular concentration. The probability distribution data for higher bi-analyte concentration and other details for ultra-low concentration measurements can be found in Supplementary Figure-7 and Note-3.

**Sample preparation for trapping experiment:**

For nanoparticles trapping experiments two metal films (gold and silver) and three nanoparticles (Ag, Ag@Au and Au) were used for comparative study. 80 µL colloidal solutions were dropped on to the metal film to observe the trapping phenomena. The colloids were diluted in water such that the number of nanoparticles/mL remained almost same for every colloidal solutions used in different experiments. The metal film thickness was kept constant to 40nm for both silver and gold films. The LASER power was kept 37mW before entering the prism throughout all trapping experiments. Videos / images were collected using 4x objective lens (NA = 0.13).

 Acknowledgements

This research work was partially funded by DST-SERB Grant (SR/S2/LOP-0007/2011) and DST Nanoscience Unit Grant (SR/NM/NS-42/2009), India. G.V.P.K. thanks DST, India for Ramanujan Fellowship.

**Author Contribution**

P.P.P. and R.C. performed all the experiments and simulations, and prepared all the figures and movies; R.P.T. assisted in sample preparation; A.D. assisted in optics experiments and simulations; G.V.P.K. conceived the idea, supervised the research and wrote the paper.

**Additional information**
Supplementary Information accompanies this paper at http://www.nature.com/naturecommunications

**Competing financial interests**: The authors declare no competing financial interests.

**Supplementary Figures:**

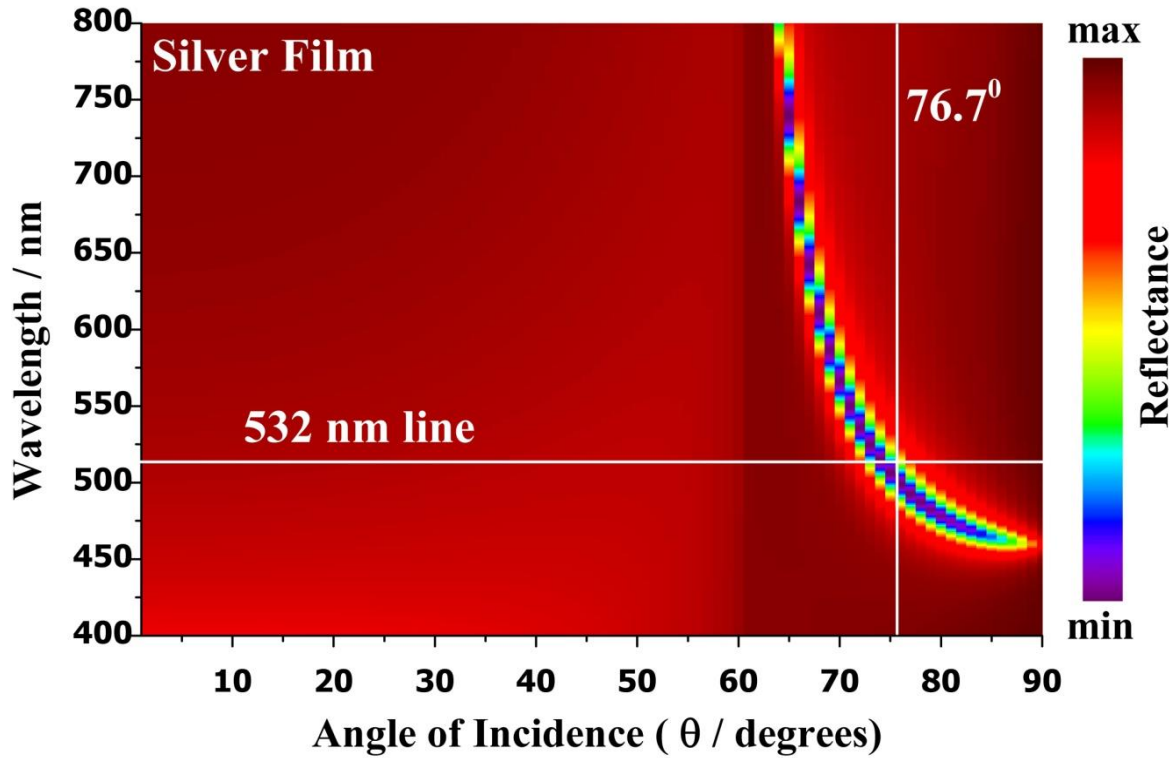

**Supplementary Figure 1. Angle dependent absorption of thin metal film:**
The angle dependent absorptions of 40nm silver film, calculated for the wavelength range from 400nm to 800nm. For silver film, at fixed wavelength 532nm (indicated by a horizontal white line) which is used for our study, the absorption maximum is around $76^0$ (indicated by a vertical white line). Calculations are based on solving Fresnel's equation for TM waves in Kretschmann configuration. MATLAB code given by Etchegoin et al.[1] was implemented for calculation. 40nm thick silver film was placed on N-BK7- glass-dove prism (RI = 1.519) and water (RI=1.33) was the outer environment.

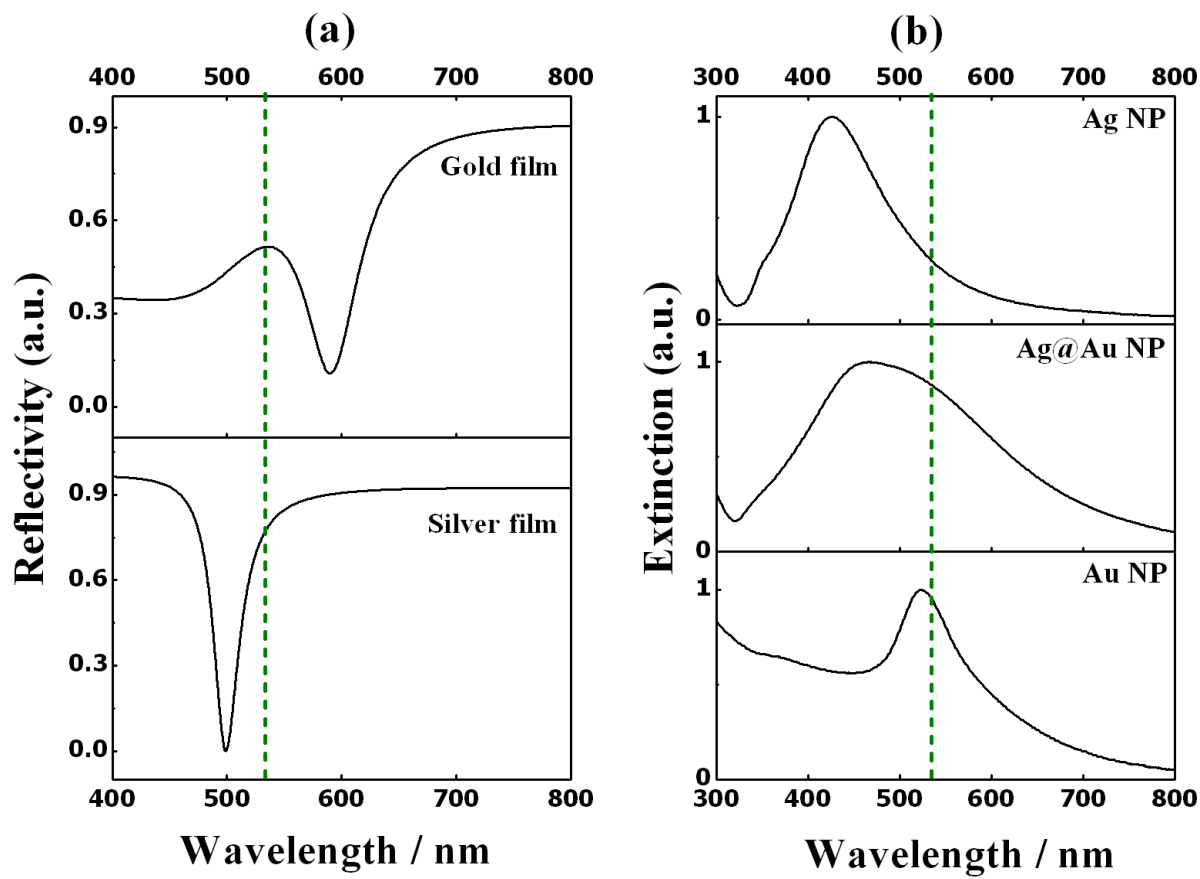

**Supplementary Figure 2. Extinction spectra of nanoparticles and thin films:** (a) Wavelength dependent reflectivity spectra were calculated for 40nm silver and gold films. (b) Normalized extinction spectra for (upper) silver NPs, (middle) Ag@Au NPs and (lower) Au NPs were measured using UV-Vis absorption spectrometer.

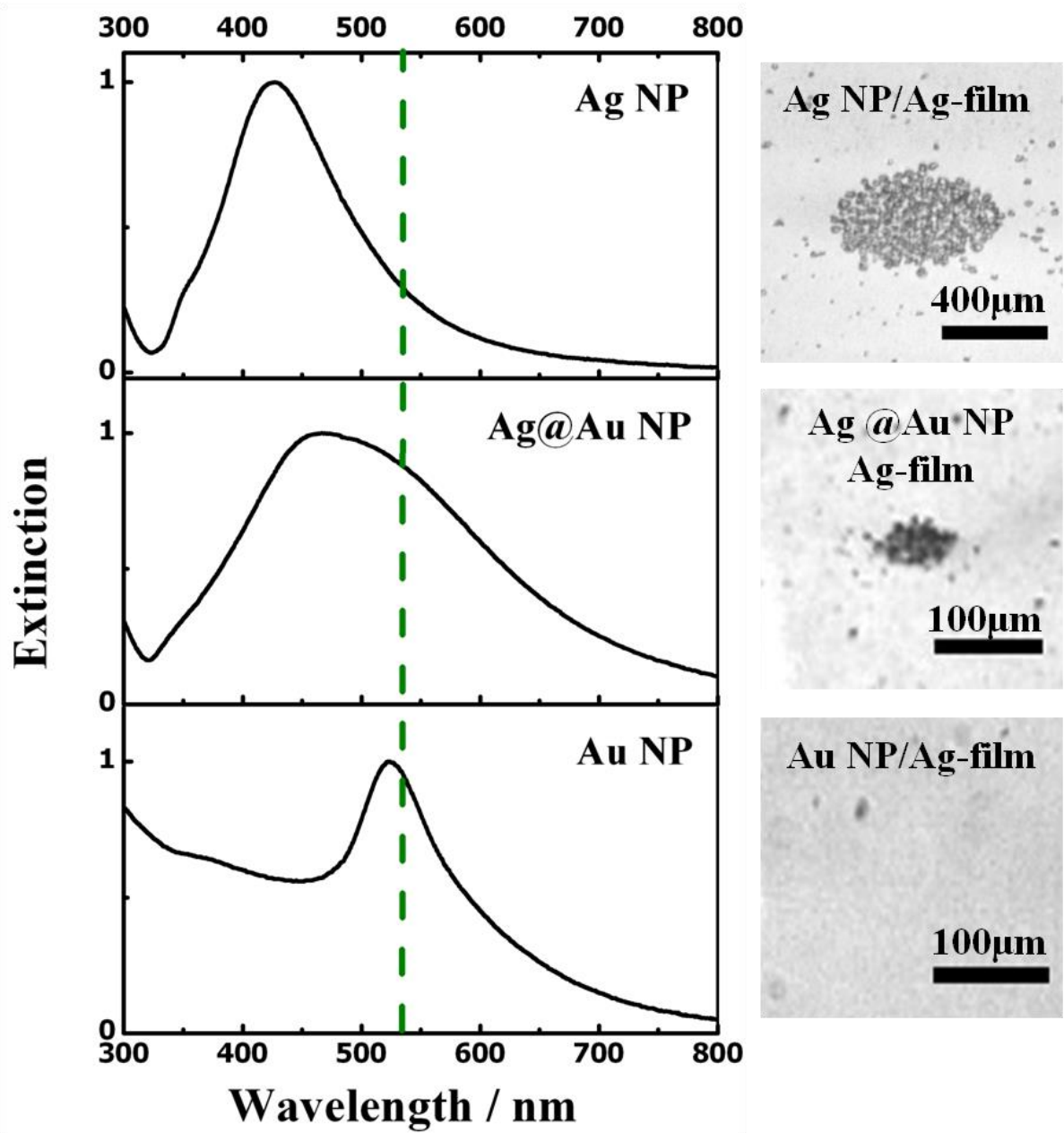

**Supplementary Figure 3. Comparison of trap sizes of different nanoparticles:** Silver nanoparticles having LSP resonance far from the excitation wavelength give rise to a larger trap size, whereas core-shell particles having LSP resonance partially overlapping with excitation wavelength result moderate trap size. In contrast to this, gold nanoparticles won't get trapped since their LSP resonance overlaps with excitation wavelength.

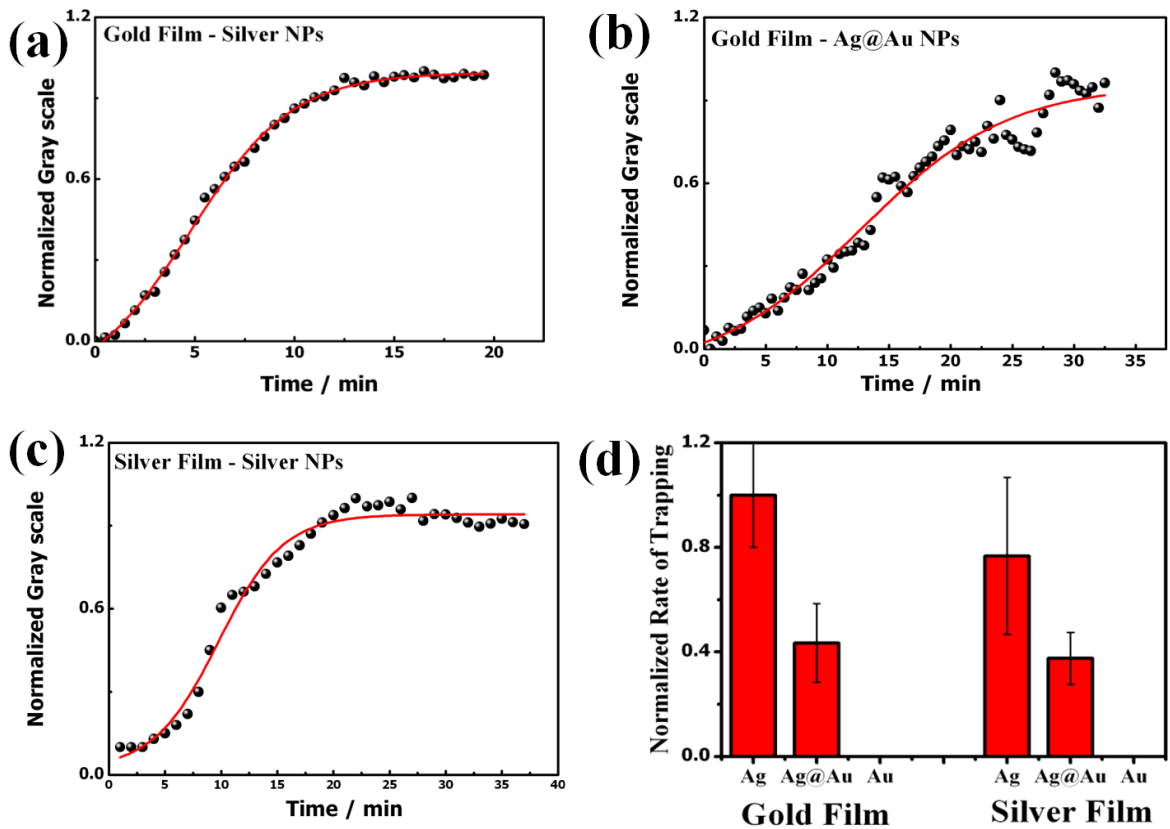

**Supplementary Figure 4. Comparison of trapping kinetics for different nanoparticles on different films:** Variation of nanoparticle density in optical potential was monitored as function of time. Particle density was calculated by measuring the grey value from the optical images captured at different time. (a) Time dependent trapping of Ag NPs on gold thin film; experimental data points (solid black spheres) have been fitted with Boltzmann curve (solid red line) to calculate the time required for trapping and rate of trapping. Similar measurements were done for (b) Ag@AuNPs on Au film and(c) Ag NPs on Ag film. We observed experimentally that Au NPs were NOT trapped either on gold film or on silver film. (d) Graph comparing the rate of trapping for different nanoparticles on different films. Error bars represent the fluctuations in the trapping rate throughout the experiment.
4

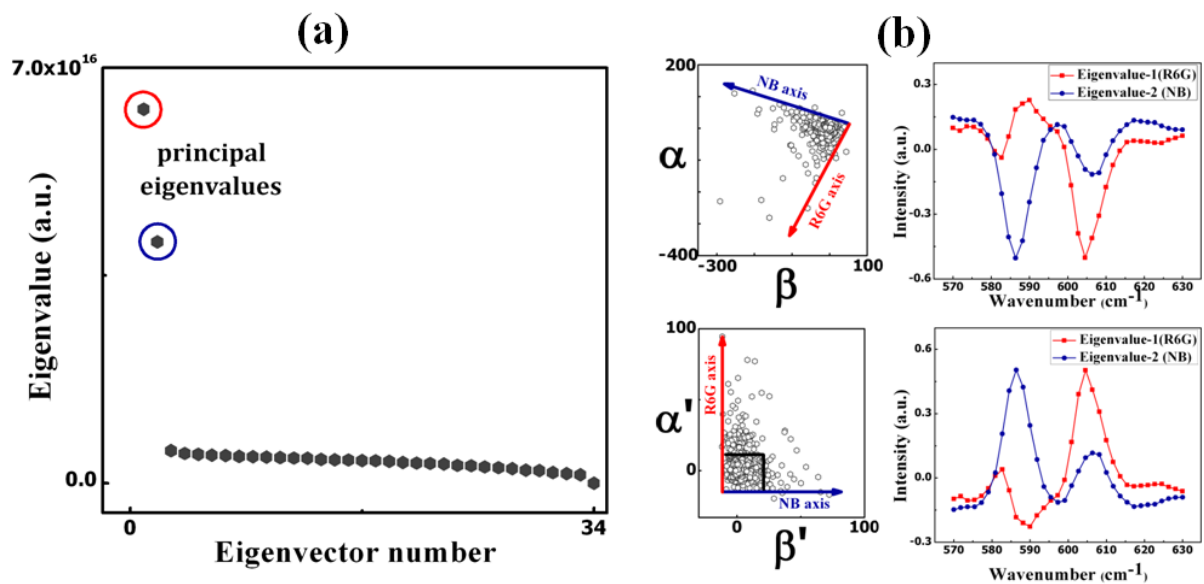

**Supplementary Figure 5. Modified Principal Component Analysis:** (a) The plot of eigenvalues against the eigenvector numbers. Here, the two principal eigenvalues denote the two dye signatures (red circle for R6G and blue for NB). (b) The coefficient matrix plot, eigenvectors against the wavenumber plot and their 2-D transformations are shown in this portion. The coefficient matrix denotes the distribution of the dye signals which are distinguished from each other and the transformed matrix shows clearly that two dye axes are orthogonal to each other. The eigenvectors are actually the spectral features of the input data matrix.

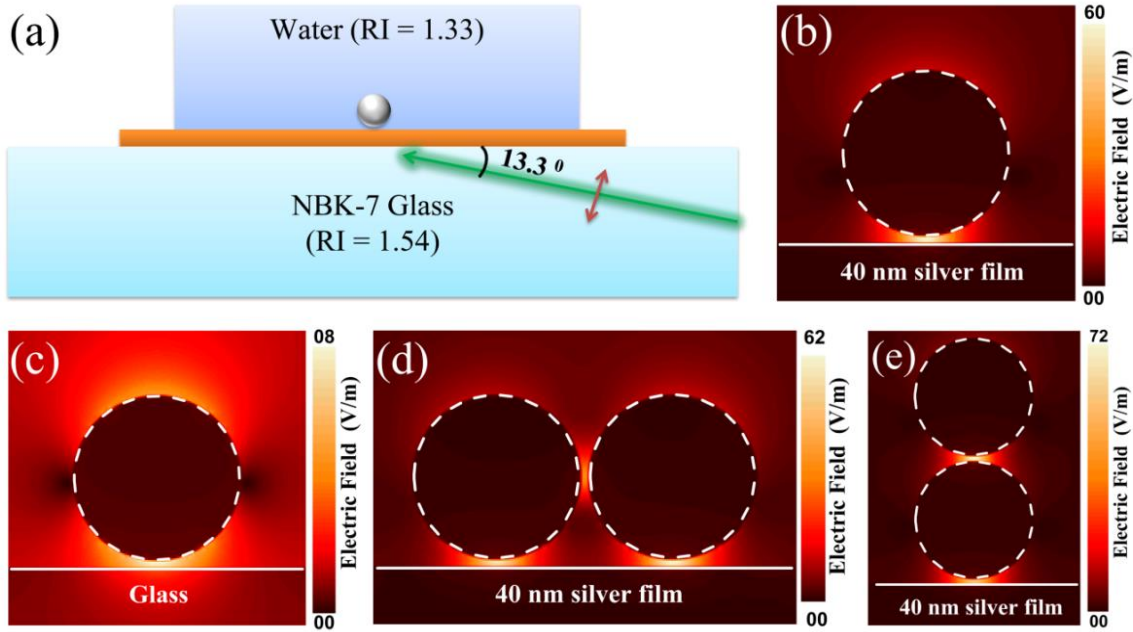

**Supplementary Figure 6. Finite domain time difference (FDTD) calculations:** (a) 3D – FDTD simulation configuration: Green line is a p-polarized 532nm Gaussian profile laser incident from a far-field regime through N-BK7 glass medium (RI = 1.519) at an angle of incidence, 76.7 degree. Grey coloured sphere is a silver-core gold-shell nanoparticle in water medium (RI = 1.33) placed 2nm above the 40nm thick silver metal (orange box). (b) Electric field distribution in the plane containing core-shell nanoparticle and silver film; film-particle junction shows maximum electric field concentration of 60 V/m; (c) Electric field distribution for a core-shell particle placed 2nm above on the glass film. (d) Electric field map for two nanoparticles clustered horizontally on the metal film with 2nm gap between the particles. Film-particle junction has higher field electric field concentration compared to particle-particle junction. (e) Electric field distribution for two core-shell nanoparticles clustered vertically on the metal film with 2nm gap between the particles; film-particle gap was also 2nm. Particle–particle junction has higher field concentration compared to particle–film junction.

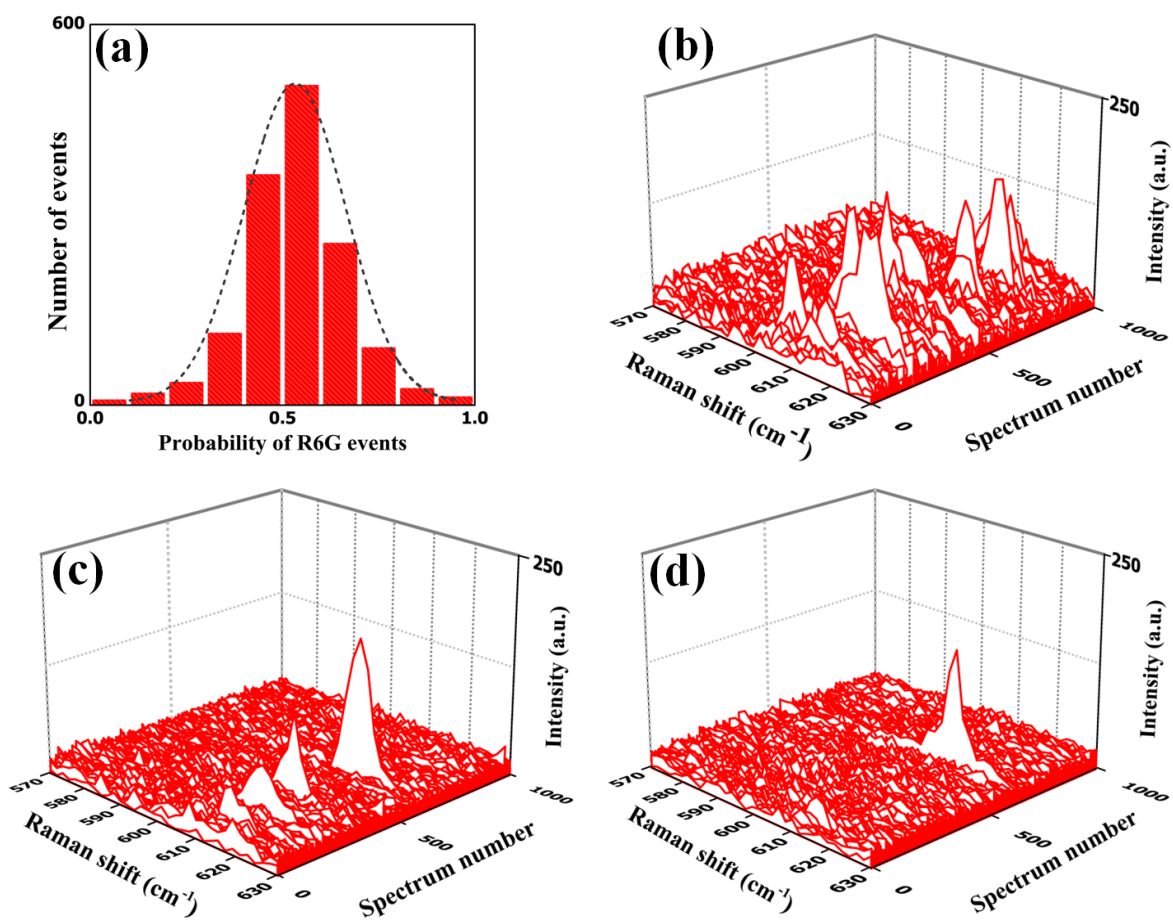

**Supplementary Figure 7. Concentration dependent study:** (a) Probability histogram of the signals acquired from higher bi-analyte dye concentration (200nM). In this case, events with probability around 0.5 were more likely to occur which means at higher dye-concentration, events due to both NB and R6G signals are dominant. Here we used 200nM concentration for both the dyes, which implies that ~960 molecules of each analyte reside per one nanoparticle.1419 events are extracted from collected total 3000 signals using MPCA. The above histogram can be contrasted to the extreme probability distribution in Figure-3 of the main manuscript. (b), (c) and (d) are time series surface-enhanced resonance Raman spectra (SERRS) of R6G at 3 different concentrations.(b) Signal fluctuation at 50nM R6G concentration (i.e. ~240 molecules of each dye per one nanoparticle). Here most of the events give the dye signal. (c) At 10 pM dye concentration (i.e. one molecule of each dye per ~200 nanoparticles) number of signals decreased as the number ratio went down by order of three. (d) The signal fluctuation at ultra-low concentration (10fM) of the analyte. Here the number ratio is like, one molecule of each dye per ~2 x $10^5$ nanoparticles, still we could detect the molecule but with a lower probability.

**Supplementary Notes:**

# 1. Growth kinetics – Boltzmann equation for the assembly of Ag@Au nanoparticles in plasmofluidic field

$$y = \frac{A_1 - A_2}{1 + e^{(t-t_0)/dt}} + A2$$

$A_1$, $A_2$ define the initial and final value of $y$ (i.e. actually the grey scale intensity). $t_0$ is the value of $t$(time) at ($y = (A_1+A_2)/2$), which can be extracted from the fit of the experimental data points. Rate of trapping was calculated by measuring the slope at $t_0$ which is given by

$$rate\ of\ trapping = \frac{(A_2 - A_1)}{4dt}$$

There are three different phases involved in this equation:

1) *Lag Phase*: Value of $y$ is almost constant during initial time period ($t$). During this phase, the number of nanoparticles in the trap is almost zero. This is due to the fact that finite amount of time is needed for convection to start, which brings the nanoparticles into the laser illuminated region.

2) *Exponential Growth Phase*: Value of $y$ grows exponentially with time $t$. During this phase, the number of the nanoparticles within the illuminated region increases exponentially.

3) *Saturation Phase*: Value of $y$ reaches to a saturation point and doesn't change with time $t$. During this phase, the number of the nanoparticles remains almost constant.

# 2. Finite-difference time-domain (FDTD) calculations:

The electromagnetic response of metal nanostructures has been simulated by 3-dimentional finite difference time domain (FDTD) calculations using commercially available software package (Lumerical Solutions, Inc.). Supplementary Figure-6 shows the simulation configuration for FDTD calculations. The metal nanosphere was made of as Ag@Au (45nm silver-core and 5nm gold-shell) nanoparticle and kept 2nm above on the 40nm thick silver metal film. The dielectric constant of silver and gold was taken from Johnson and Christy[2]. A p-polarized 532nm Gaussian-profile laser beam was incident from a far-field regime through N-BK7 glass medium (RI = 1.519) at an angle of $76.7^0$ to create an evanescent-field excitation geometry. The fluidic environment of the Ag@Au NP was kept as water medium (RI = 1.33) to mimic the experimental setup. The electric field distribution in the plane of particle and film was calculated with the mesh size of $0.5\times0.5\times0.5$ nm$^3$. 52,500 iteration time-steps were used to ensure the stability of the calculated solutions.

## 3. Detection at ultra-low concentrations

To test the detection limits of our method, we performed time-series SERS experiments with R6G at 3 different concentrations: 1 fM, 10pM and 50nM, as in the supplementary figure 7 (b), (c) and (d). We observed that if we go down to the lower concentration, the SERS intensity fluctuates and importantly, the occurrence of the signals decreases drastically.

## Supplementary References:

Movie links can be found in the below link:

https://media.nature.com/original/nature-assets/ncomms/2014/140707/ncomms5357/extref/ncomms5357-s2.mov

https://media.nature.com/original/nature-assets/ncomms/2014/140707/ncomms5357/extref/ncomms5357-s3.mov

https://media.nature.com/original/nature-assets/ncomms/2014/140707/ncomms5357/extref/ncomms5357-s4.mov

https://media.nature.com/original/nature-assets/ncomms/2014/140707/ncomms5357/extref/ncomms5357-s5.mov